\documentstyle[epsf,11pt,amssymbols]{article}

\newcommand{\xo}{x_o}
\newcommand{\phio}{\phi_{o}}
\newcommand{\ao}{a_{o}}
\newcommand{\OandT}{\"{O}zer and Taha}

\begin{document}

\thispagestyle{empty}%

{\raggedleft
Preprint: NZ-CAN-RE-93/2 \par
PACS: 0420 1210 9880\par
}

\vfill

	\begin{center}
		{\Large\bf\expandafter{Exact superstring  motivated\\
cosmological models}\par}
	\end{center}
	\vfill
	\begin{center}
		 {\Large Richard Easther \\}
                  \vspace{10mm}
                 { Department of Physics and Astronomy \\
                   University of Canterbury \\
                   Private Bag 4800 \\
                   Christchurch \\
                   New Zealand. \\}
        \end{center}
        \vspace{5mm}
        \begin{center}
                   email: r.easther@phys.canterbury.ac.nz \\
	\end{center}

\vfill
\mbox{}
\newpage

\setcounter{page}{1}
\section*{Abstract}
We present a number of new, exact scalar field cosmologies where the
potential consists of two or more exponential terms. Such potentials are
motivated by supergravity or superstring models formulated in higher
dimensional spacetimes that have been compactified to (3+1)-dimensions.

We have found solutions in both curved and flat Robertson Walker
spacetimes. These models have a diverse range of properties and often
possess several distinct phases, with a smooth transition between
ordinary and inflationary expansion. While exponential potentials
typically produce powerlaw inflation, we find models where the
inflationary period contains eras of both powerlaw and exponential
growth.

\vspace{1cm}

\section{Introduction}

Since the first inflationary cosmological models were proposed
\cite{Guth1981a,AlbrechtET1982a,Linde1982a} most calculations have been
performed using one of a variety of approximations. While these are
usually sufficient and often unavoidable, there is considerable interest
in scalar field cosmologies that can be solved exactly.

In this paper we introduce a number of exact solutions for
Robertson Walker universes that contain a single scalar field, $\phi$,
with the potential,
\begin{equation}
V(\phi) = \sum_{j=1}^{N}{\Lambda_j \exp{(-\lambda_j\phi)}}.
\label{Vgeneral}
\end{equation}
Potentials of this type characteristically arise when a
higher-dimensional theory is compactified to $(3+1)$-spacetime,
including various supergravity and superstring models
\cite{%
SalamET1984a,%
FradkinET1985a,%
CallanET1985a,%
CallanET1986a,%
GrossET1987a,%
Halliwell1987a,%
CampbellET1991a}.
In particular, as \OandT\ \cite{OzerET1992a} point out, the potential
\begin{equation}
V(\phi) = \sum_{j=1}^{N}{\Lambda_j \exp{(-j\gamma\phi)}},
\label{Vgenerala}
\end{equation}
where $\gamma$ is a constant, is motivated by a perturbation expansion
in superstring theories. In scalar field cosmologies with exponential
potentials $\gamma$ is usually a free parameter and in this paper we
discuss solutions with a variety of different values of
$\gamma$. However, string theory makes the generic prediction that
$\gamma = \sqrt{2}$ which must be incorporated into any fully realistic
theory.  Cosmological models based directly upon the superstring action
have been examined by Casas, Garc\'{\i}a-Bellido and Quir\'{o}s
\cite{CasasET1991b,GarciaBellidoET1992a}. Brustein and Steinhardt
\cite{BrusteinET1993a}  demonstrate that there are severe difficulties in
implementing a realistic cosmology based on the conventional formulation
of superstring theories, but this is not being attempted here.

Exact solutions for potentials containing a single exponential
term with  restrictions either on the initial conditions or the choice of
model parameters have been discussed by several authors
\cite{LucchinET1985a,%
BarrowET1986a,%
Barrow1987a,%
BurdET1988a,%
YokoyamaET1988a,%
Muslimov1990a,%
Liddle1989a,%
Ratra1992a}.
The full solution was obtained by Salopek and Bond \cite{SalopekET1990a}
and generalised by Lidsey to a two-field model
\cite{Lidsey1992a}. Solutions for potentials of the type
$V(\phi)=A^2e^{2\lambda\phi}+B^2e^{-2\lambda\phi}-2AB$ have been found
by de Ritis {\em et. al.\/}
\cite{DeRitisET1990b,DeRitisET1990a,DeRitisET1991a}.

Powerlaw inflation \cite{%
LucchinET1985a,Barrow1987a,Liddle1989a,%
Ratra1992a,AbbottET1984a,LucchinET1985b},
where $a(t) \propto t^p$ and $p>1$, is normally driven by a
scalar field with $V(\phi)=\Lambda e^{-\lambda\phi}$. We generalise this
potential to models where $V(\phi)$ has the form of
equation~(\ref{Vgeneral}).  In Sections 3 and 4 we give several exact
solutions in curved spacetime for a potential which has two exponential
terms.  The flat spacetime case is discussed in Section 5 and here
we find a diverse range of models. For instance, examples where
powerlaw growth is preceeded by a period of exponential expansion are
presented, as well as models where powerlaw inflation commences after a
finite amount of non-inflationary growth.

Like virtually all exact solutions, those presented here must be viewed
as toy models rather than attempts to construct a realistic
cosmology. In this sense, the inflationary solutions in this paper are
analagous to intermediate inflation \cite{Barrow1990a,BarrowET1990c}
which is based on an unlikely looking potential but has received
attention because it can be solved exactly.  Also, by studying exact
solutions which widen the repertoire of possible inflationary behaviour
we create new possibilities which can then be incorporated into more
complicated models.

\section{The Equations}

In this section we present the equations which govern the evolution of a
Robertson Walker universe containing a single scalar field, $\phi$, with
effective potential $V(\phi)$.

The density, $\rho$, is
\begin{equation}
\rho = \frac{1}{2}\dot{\phi}^2 + V(\phi). \label{rhodef}
\end{equation}
Expressed in natural units, the Einstein field equations for a
Robertson Walker universe, are \cite{KolbBK1,LindeBK1}
\begin{eqnarray}
H^2 &=& \frac{\rho}{3} - \frac{k}{a^2}, \label{Hdef} \\
\dot{H} &=& -\frac{1}{2}\dot{\phi}^2 + \frac{k}{a^2}, \label{Hdotdef}
\end{eqnarray}
where $a(t)$ is the scale factor and $H = \dot{a}/a$ is the Hubble
parameter. The semi-classical equation of motion for $\phi$ is
\begin{equation}
\ddot{\phi} + 3H\dot{\phi} + \frac{dV(\phi)}{d\phi} = 0. \label{eofmotion}
\end{equation}
Differentiating $\rho$ and comparing the result with
equation~(\ref{eofmotion}) gives
\begin{equation}
\dot{\rho} = -3H\dot{\phi}^2. \label{rhodot}
\end{equation}
While taking the time, $t$, as the independent variable is the most
natural choice, the mathematical complexity of exact solutions to
equations~(\ref{rhodef}) to~(\ref{eofmotion}) is often reduced by
parametrising the motion in terms of the field, $\phi$. This formalism
was introduced by Muslimov, Salopek and Bond
\cite{Muslimov1990a,SalopekET1990a} but the form of it used here is due
to Lidsey~\cite{Lidsey1991b}. In this paper, we will find it useful to
make the extra substitution,
\begin{equation}
x = \exp{(-\xi\phi)}, \label{xdef}
\end{equation}
where $\xi$ is a constant.

The shift between $x$ and $t$ is obtained by rearranging
equation~(\ref{rhodot})
\begin{equation}
\frac{dx}{dt} = -\frac{\xi^2x^2}{3H}\frac{d\rho}{dx}. \label{drhodx}
\end{equation}
Employing equation~(\ref{drhodx}) we find a set of equations, expressing
$\rho$, $V$, $a$, $t$ and $H$ as functions of $x$. From
equation~(\ref{rhodot}) it follows that \cite{Lidsey1991b}
\begin{equation}
6H^2 = -\xi^2 x^2 \frac{\rho'\chi'}{\chi} \label{Hxdef}
\end{equation}
where $\chi = a^2$ and the dash denotes differentiation with respect to $x$.
Comparing equations (\ref{Hdef}) and (\ref{Hxdef}) yields a first order
differential equation for $\chi$,
\begin{equation}
\xi^2x^2\rho'\chi' + 2\rho\chi = 6k.  \label{chide}
\end{equation}
The general solution of this equation is
\begin{equation}
\chi(x) =
\exp{\left[-\frac{2}{\xi^2}\int_{x_0}^{x}\frac{\rho}{x^2\rho'}dx\right]}
\left\{a_{0}^2 +\frac{6k}{\xi^2}\int_{x_0}^x
\exp{\left[\frac{2}{\xi^2}
\int_{x_0}^{x}\frac{\rho}{x^2\rho'}dx\right]}
\frac{dx}{x^2\rho'}\right\}, \label{chisol}
\end{equation}
where $\ao$ and $\xo$ and the values of $a$ and $x$ when $t=0$. For any given
density, $\rho$, the potential is
\begin{equation}
V(x) = \rho -\frac{(\xi x \rho')^2}{18H^2}
\end{equation}
or, equivalently,
\begin{equation}
V(x) = \rho + \frac{\chi\rho'}{3\chi'}. \label{Vfromrho}
\end{equation}
The time, $t(x)$, is calculated by integrating equation~(\ref{drhodx}),
\begin{equation}
t(x) = -\frac{3}{\xi^2}\int_{\xo}^x \frac{H}{x^2\rho'}dx. \label{timex}
\end{equation}
Finally $\Omega$, the ratio between the density and the
critical density,  is  \cite{MisnerBK1}
\begin{equation}
\Omega(x) = \frac{\rho(x)}{3H^2(x)}.  \label{Omegax}
\end{equation}

Not all exact scalar field cosmologies are inflationary. During
inflation the comoving coordinate volume contained inside the horizon
decreases as the universe expands or, equivalently,  $\ddot{a} > 0$
\cite{AbbottET1984a}. In terms of the parameter $\epsilon =
-\dot{H}/H^2$,  inflation is taking place when $\epsilon < 1$. Using
equation~(\ref{drhodx}) we write $\epsilon$ in terms of $x$,
\begin{equation}
\epsilon(x) = \xi^2x^2\frac{\rho'H'}{3H^3} \label{epsilondef}
\end{equation}

The method we use to find exact solutions is to specify a particular
form of the density, $\rho(x)$, and then run the field equations
backwards to derive the potential that produced it. The choice of
$\rho(x)$ is guided by our interest in potentials that are of the
general form, equation~(\ref{Vgeneral}). However, as long as the
integrals in equations~(\ref{chisol}) and (\ref{Vfromrho}) can be
performed then this method will work for an arbitrary $\rho(x)$. While
the exact solutions presented here could have been written down without
any comment as to how they were found, it will hopefully be helpful to
the reader if the methodology is made explicit.

Other authors have obtained exact solutions by constraining the field
equations in some way and then solving the restricted problem.  In
particular, Ellis and Madsen \cite{EllisET1991a} derive several exact
solutions by specifying the functional form of the scale factor, $a(t)$,
and then computing $V(\phi)$. While the calculation involved here is
similar, they emphasise the desired expansion whereas we are seeking
exact solutions when the potential is of the generic form,
equation~(\ref{Vgeneral}).

Virtually all the exact solutions that are to be found in the literature
do not hold over the complete range of initial conditions, and those
discussed in this paper are no exception.  Also, by characterising the
motion with the field, $\phi$ we are implicitly assuming that the $\phi$
is strictly increasing or decreasing. When this assumption breaks down,
we can describe the solution in a piecewise way. An example of this is
given in Section 4.

\section{Positively Curved Spacetime}

We begin by considering models in Robertson Walker spacetimes
with positive curvature ($k=1$).  \OandT\ \cite{OzerET1992a}
give two exact solutions for the potential
\begin{equation}
V(x) = Cx - Dx^2, \quad \xi = 1, \label{Vpos}
\end{equation}
which they designate Solutions I and II. While their solutions are
distinct from one another when the time is chosen as the independent
variable, they both have the density,
\begin{equation}
\rho(x) = Ax - Bx^2, \label{rhopos}
\end{equation}
where $A$ and $B$ depend on the initial conditions. By substituting
$\rho(x)$ into equations~(\ref{chisol}) and~(\ref{Vfromrho}) we derive
the most general potential that produces a density of the form
equation~(\ref{rhopos}),
\begin{equation}
V(x) = \frac{5A\xo(A\ao^2\xo - 6)x +
4(3A - 2AB\ao^2\xo^2 + 6B\xo)x^2 + 2B(B\ao^2\xo^2 -3)x^3}
{6\xo(A\ao^2\xo - 6) + 6(3- B\ao^2\xo^2)x}. \label{trialV}
\end{equation}
We can carry out this procedure for other choices of $\rho(x)$ and
$\xi$. However, this case is special in that for other relatively simple
forms of $\rho(x)$ in curved spacetime the corresponding potential is
extremely complicated.

When $A=6/\ao^2\xo$ or $B=3/\ao^2\xo^2$ the potential,
equation~(\ref{trialV}), simplifies to the form of equation~(\ref{Vpos}).
We consider these two special cases in turn. First, setting $A =
6/\ao^2\xo$ in equation~(\ref{rhopos}) and using
equations~(\ref{chisol}) and (\ref{timex}) gives the following solution,
parametrised by $x$:
\begin{eqnarray}
V(x) &=& \frac{2}{3}Ax - \frac{1}{3}Bx^2 \label{VAx}, \\
 H^2(x) &=&
\frac{A}{6}x - Dx^2, \label{H2Ax} \\
a(x) &=& \ao\sqrt{\frac{\xo}{x}},
\label{aAx} \\
t(x) &=& \ao\left[\sqrt{ \frac{\xo}{x}- D\ao^2\xo^2} -
\sqrt{1 - D\ao^2\xo^2}\right], \label{tAx}
\end{eqnarray}
where we have identified $B=3D$ from equations~(\ref{Vpos}) and (\ref{VAx}).
Inverting gives $a(t)$ and $\phi(t)$ and generalises Solution I of
\OandT.
\begin{eqnarray}
\phi(t) &=& \phio + \log{\left[D\ao^2\xo^2 +
\left(\frac{t}{\ao} + \sqrt{1- D\ao^2\xo^2}\right)^2\right]},
\label{phiAt} \\
a(t) &=& \ao\sqrt{D\ao^2\xo^2 +
\left(\frac{t}{\ao} + \sqrt{1 -  D\ao^2\xo^2}\right)^2}.
\label{aAt}
\end{eqnarray}
The only constraint is the requirement that $H^2 \geq 0$ when $x = \xo$,
or $D\leq 1/\ao^2\xo$.  Setting $D = 1/\ao^2\xo$ recovers \OandT's
result. For this solution $\epsilon(x)$, equation~(\ref{epsilondef}), is
\begin{equation}
\epsilon(x) = \frac{1-2D\ao^2\xo x}{1-D\ao^2\xo x}
\end{equation}
When $D>0$, $a(t) \neq 0$ and $\epsilon < 1$ so this solution is always
inflationary and non-singular, as \OandT\ \cite{OzerET1992a} point out.
At large negative times, $a(t)$ is decreasing towards a finite minimum
size, after which it will expand indefinitely. While this is an
inflationary solution, it is not asymptotically flat, since
\begin{equation}
\Omega = \frac{2 - D \ao^2 \xo x}{1 - D\ao^2 \xo x}
\end{equation}
and $\Omega \rightarrow 2$ at small $x$, or large times. Further
examples of inflationary models where $\Omega \neq 1$ can be found in
Ellis {\em et. al.\/} \cite{EllisET1991c}. If $D < 0$ then $a(t) = 0$ at
some finite, negative time and the resulting model universe does contain
an initial singularity.  In this case inflation never begins,
since $\epsilon > 1$ at all times.

Now put $B = 3/\ao^2\xo^2$, giving the solution
\begin{eqnarray}
V(x) &=& \frac{5}{6}Ax - \frac{2}{3}Bx^2 \label{VBx}, \\
H^2(x) &=& \frac{2}{5}Cx - \frac{2}{3}Bx^2, \label{H2Bx} \\
\phi(t) &=& \phio + \log{\left[\frac{5}{C\ao^2\xo} +
\frac{C\xo}{5}\left(\frac{t}{\sqrt{2}} +
\frac{5}{C}\sqrt{\frac{C}{5\xo} -
\frac{1}{\ao^2\xo^2}}\right)^2\right]}, \label{phiBt} \\
a(t) &=& \ao\left[\frac{5}{C\ao^2\xo} +
\frac{C\xo}{5}\left(\frac{t}{\sqrt{2}} +
\frac{5}{C}\sqrt{\frac{C}{5\xo} -
\frac{1}{\ao^2\xo^2}}\right)^2   \right] \label{aBt}
\end{eqnarray}
where $C = 5A/6$. To ensure that $H^2 \geq 0$, we need $C\geq
5/\ao^2\xo$.  Setting $C = 5/\ao^2\xo$ recovers Solution II of \OandT.
This is a non-singular model and at large times $a(t) \propto
t^2$ and $\Omega \rightarrow 1$, so the solution approaches powerlaw
inflation in flat spacetime.

\section{Negatively Curved Spacetime}

The next possibility we consider is the existence of exact models
in a Robertson Walker universe with negative curvature ($k=-1$). We look
for the potential that gives a density of the form
\begin{equation}
\rho(x) = Ax + Bx^2, \quad \xi = 1. \label{rhoneg}
\end{equation}
{}From equations~(\ref{chisol}),~(\ref{Vfromrho}) and (\ref{rhoneg}) we obtain
\begin{equation}
V(x) = \frac{5A\xo(A\ao^2\xo + 6)x +
4(3A + 2AB\ao^2\xo^2 + 6B\xo)x^2 + 2B(B\ao^2\xo^2 -3)x^3}
{6\xo(A\ao^2\xo + 6) + 6(3- B\ao^2\xo^2)x}. \label{trialVneg}
\end{equation}
Choosing either $A=-6/\ao^2\xo$ or $B=3/\ao^2\xo^2$ simplifies the
potential, equation~(\ref{trialVneg}), to the generic form given by
equation~(\ref{Vgeneral}). As is the case for positively curved
spacetime, other choices of $\rho(x)$ result in a very complicated
potential.

Putting $B=3/\ao^2\xo^2$ gives the solution
\begin{eqnarray}
V(x) &=& \frac{5}{6}Ax + \frac{2}{3}Bx^2, \label{VnegBx} \\
H^2(x) &=& \frac{A}{3}x + \frac{2}{3}Bx^2, \label{H2negBx} \\
\phi(t) &=& \phio + \log{\left[ \frac{A\xo}{6}\left(\frac{t}{\sqrt{2}}
+ \frac{A}{6}\sqrt{\frac{1}{\ao^2\xo^2} + \frac{A}{6\xo}}\right)^2
- \frac{6}{A\ao^2\xo^2}\right]}, \label{phiBtneg} \\
a(t) &=& \ao\left[ \frac{A\xo}{6}\left(\frac{t}{\sqrt{2}}
+ \frac{6}{A}\sqrt{\frac{1}{\ao^2\xo^2} + \frac{A}{6\xo}}\right)^2
- \frac{6}{A\ao^2\xo^2}\right]. \label{atnegB}
\end{eqnarray}
This universe begins with an initial singularity ($a=0$) and expands
forever. It is always inflationary and at large times $\Omega
\rightarrow 1$, so it is asymptotically flat.

When $A = -6/\ao^2\xo$ the solution is markedly different from those
others we have examined so far. The scale factor $a(t)$ initially
increases but the expansion comes to a halt at a finite time, after
which the universe contracts towards $a(t)=0$. We have to patch two
solutions together, one for the expanding phase, and one for the
contracting phase, with the Hubble parameter, $H$, having the
appropriate sign in equation~(\ref{timex}). Defining $B = 3D$, we find
\begin{eqnarray}
V(x) &=& \frac{2}{3}A + \frac{B}{3}x^2, \label{VnegAx} \\
H^2(x) &=& \frac{A}{6}x + Dx^2, \label{H2negAx} \\
a(x) &=& \ao\sqrt{\frac{\xo}{x}}, \label{anegAx} \\
t(x) &=& \ao\left[ \sqrt{D\ao^2\xo^2 - 1} - \sqrt{D\ao^2\xo^2 -
\frac{\xo}{x}}\right], \quad H > 0. \label{tnegAx1}
\end{eqnarray}
If $x < x_c$, where $x_c = 1/D\ao^2\xo$, then $t(x)$ acquires a
complex portion. However $H(x_c) = 0$ and $a(x_c)$ is the maximum value of
$a(x)$. So when $t > t(x_c)$, $H < 0$ and  the universe is contracting. In
this case $V(x), H^2(x)$ and $a(x)$ are the same as those calculated
above but
\begin{equation}
t(x) - t_1 = a_1\left[\sqrt{Da_1^2x_1^2 - \frac{x_1}{x}} -
\sqrt{Da_1^2x_1^2 -1}\right], \quad H < 0. \label{tnegAx2}
\end{equation}
Choosing the initial conditions in equation~(\ref{tnegAx2}) to be $a_1 =
a(x_c)$, $t_1 = t(x_c)$ and $x_1 = x_c$ gives $t(x)$ during the
contraction phase of the universe described by equations~(\ref{VnegAx})
to (\ref{anegAx}),
\begin{equation}
t(x) = \ao\left[\sqrt{D\ao^2\xo^2 -1} + \sqrt{D\ao^2\xo^2 -
\frac{\xo}{x}}\right], \quad  H < 0. \label{tnegAx3}
\end{equation}
These equations may now be inverted, giving
\begin{eqnarray}
\phi(t) &=& \phio + \log{\left[D\ao^2\xo^2 -
\left(\frac{t}{\ao} - \sqrt{D\ao^2\xo^2 -1}\right)^2 \right]} ,
\label{phinegAt} \\
a(t) &=& \ao\sqrt{D\ao^2\xo^2 - \left(\frac{t}{\ao} -
\sqrt{D\ao^2\xo^2 -1}\right)^2}. \label{anegAt}
\end{eqnarray}

This is not a viable inflationary model. However, it is an example of an
exact scalar field cosmology in a curved spacetime and it serves as a
reminder that not all exact solutions will be inflationary.

\section{Flat Spacetime}

In curved spacetime, we could only find a few exact solutions where the
potential, $V(x)$, had a relatively simple form and there was typically
some connection between the allowable range of initial conditions and
one of the coefficients in the potential. In flat spacetime, however,
$k=0$ and $\rho = 3H^2$ so equations~(\ref{chisol}) to (\ref{timex}) and
(\ref{epsilondef}) become
\begin{eqnarray}
a(x) &=&
\ao\exp{\left[-\frac{1}{2\xi^2}\int_{\xo}^{x}\frac{H}{x^2H'}dx\right]}
\label{axflat} \\
V(x) &=& 3H^2 -2\xi^2x^2H'^2 \label{Vxflat} \\
t(x) &=& -\frac{1}{2\xi^2}\int_{\xo}^x \frac{1}{x^2H'}dx
\label{txflat} \\
\epsilon(x) &=& 2\left(\xi x \frac{H'}{H}\right)^2. \label{epsilonflat}
\end{eqnarray}
When $k=0$ it is simpler to specify $H(x)$ rather than $\rho(x)$.  Any
$H(x)$ that is a polynomial in $x$ will, upon substitution into
equation~(\ref{Vxflat}), give a potential that is automatically of the
form equation~(\ref{Vgeneral}) so we can find any number of exact
inflationary models in flat spacetime.

\subsection{Powerlaw Inflation}
The simplest choice of $H(x)$ is
\begin{equation}
H(x) = Ax. \label{Hxflat1}
\end{equation}
 Substituting
equation~(\ref{Hxflat1}) into equation~(\ref{Vxflat}) gives the
potential,
\begin{equation}
V(x) = (3 -2\xi^2)A^2x^2 \label{Vxflat1}
\end{equation}
The parametric solution for $a(t)$ is
\begin{eqnarray}
a(x) &=& \ao\left(\frac{\xo}{x}\right)^{1/2\xi^2}
\label{axflat1}\\
t(x) &=& \frac{1}{2A\xi^2}\left(\frac{1}{x} - \frac{1}{\xo}\right) .
\label{txflat1}
\end{eqnarray}
This solution is given by Muslimov~\cite{Muslimov1990a} but we quote it
for convenience as many of the models discussed later tend to this
solution as the time, $t$, becomes large.

Inverting, to make $t$ the independent variable,
\begin{eqnarray}
\phi(t) &=& \phio + \frac{1}{\xi}\log{\left(1 + 2A\xi^2e^{-\xi\phio}t\right)}
\label{phitflat1} \\
a(t) &=& \ao\left(1+2A\xi^2e^{-\xi\phio} t\right)^{1/2\xi^2}
\label{atflat1}
\end{eqnarray}
For large $t$, $a(t) \propto t^{p}$ where $p = 1/2\xi^2$, which is
powerlaw inflation if $\xi < \sqrt{1/2}$. While this is not the general
solution, it is an
attractor \cite{Halliwell1987b}.

\subsection{Exponential expansion from an exponential potential}
We now turn our attention to new, exact solutions.
We start with
\begin{equation}
H(x) = Ax - Bx^2, \quad B>0 \label{Hxflat2}
\end{equation}
which  results in
\begin{equation}
V(x) = A^2(3-2\xi^2)x^2 + 2AB(4\xi^2 - 3)x^3 + B^2(3-8\xi^2)x^4.
\label{Vxflat2a}
\end{equation}

{}From equations~(\ref{axflat}) to (\ref{txflat}) it follows that
\begin{eqnarray}
a(x) &=&
\ao\left[\frac{\xo^2(A-2Bx)}{x^2(A-2B\xo)}\right]^{1/4\xi^2}
\label{axflat2} \\
t(x) &=& \frac{1}{A\xi^2}\left[\frac{1}{2}
\left(\frac{1}{x}-\frac{1}{\xo}\right) +
\frac{B}{A}\log{\left(\frac{\xo(A-2Bx)}{x(A-2B\xo)}\right)}\right].
\label{txflat2}
\end{eqnarray}
This solution cannot be easily inverted, so we will work with the
parametric form. We treat this case in some detail as much of the
analysis here will apply with minor modifications to the other exact
solutions discussed in this section.

The potential $V(x)$, equation~(\ref{Vxflat2a}), possesses turning
points at
\begin{equation}
x_{a} = \frac{A}{2B}, \quad x_{b} =
\frac{A}{B}\frac{3-2\xi^2}{3-8\xi^2}. \nonumber
\end{equation}
Since $x$ is always positive, $x_b$ is only significant if it is
greater than zero.   For all values of $\xi$, $x_a$ is a local maximum
of $V(x)$. However, the potential is only bounded below when $\xi <
\sqrt{3/8}$ and  has a range of different forms, depending on the
value of~$\xi$.

Like most exact scalar field cosmologies this solution applies to a
restricted set of initial conditions, specifically $x\rightarrow x_a$ as
$t\rightarrow -\infty$, so the field is always evolving away from the
unstable equilibrium at $x_a$. Formally, $a(t) > 0$ for all negative
times and this solution lacks an initial singularity but small
perturbations in $\phi$ or $\dot{\phi}$ when $x$ is very close to $x_a$
render the solution classically unstable in this region.  However the
behaviour of the exact solution at later times is representative of a
class of models that approximate these initial conditions and we
discuss the solution in this context. If $\xo > x_a$ then $t(x)$
increases with $x$.  When $x \rightarrow \infty$ we find
\begin{equation}
t = \frac{1}{A\xi^2}\left[\log{\left(\frac{2B}{2B\xo - A}\right)} -
\frac{1}{2\xo}\right], \quad a = 0.
\end{equation}
and $a(x)$ becomes zero after a finite time. Thus this model universe
collapses into a singularity for some choices of the initial conditions.

We now focus on the case where $\xo < x_a$, and the expansion continues
indefinitely. For this solution,
\begin{equation}
\epsilon(x) = 2\xi^2\left(\frac{A-2Bx}{A-Bx}\right)^2.
\label{epsflat2}
\end{equation}
If $\xi < \sqrt{1/2}$ then $\epsilon < 1$ for all $x$ and inflation
continues forever. For larger values of $\xi$ the inflationary phase
will cease when $\epsilon =1$, or
\begin{equation}
x = \frac{A}{B} \frac{2\xi - \sqrt{2}}{4\xi -\sqrt{2}}.
\end{equation}
Furthermore, this solution has the capacity for both powerlaw and
exponential inflation. During quasi-exponential expansion, $H$ must be
relatively constant on timescales of $1/H$, during which the universe
expands by a single e-folding. This requirement is satisfied when
$\epsilon \ll 1$. So if $\xo \approx x_a$ (but not so close to the local
maximum that the solution is unstable) then $a(t)$ is approximately
exponential.

Thus this solution exhibits the properties of the two major
inflationary models at different stages in its evolution. If $\xi
\lesssim 0.2$ and $\xo \approx x_a$ then a large amount of exponential
expansion is possible, with the number of e-foldings depending
critically on $\xi$ and the ratio $\xo/x_a$. In figures~(\ref{Vphi_fig})
and (\ref{logavt_fig}) $V(\phi)$ and $a(t)$ are plotted for a
representative set of parameter values. Physically, since $x_a$ is a
local maximum, the potential is approximately flat when $x\approx x_a$.
Thus both $x$ and $V(x)$ are changing slowly, leading to an era of
quasi-exponential expansion. As $x$ moves further away from $x_a$ the
potential becomes steeper and the era of powerlaw expansion begins.

For most values of $\xi$, the coefficients of only two of the three
coefficients in $V(x)$, equation~(\ref{Vxflat2a}) are independent.
However, when $\xi = \sqrt{3/2}$, $\sqrt{3/4}$ or $\sqrt{3/8}$ one of
the terms in $V(x)$ drops out, leaving only two nonzero terms which may
be chosen arbitrarily.  In particular, when $\xi=\sqrt{3/4}$, the potential
is
\begin{equation}
V(\phi) =\frac{3}{2}A^2\exp{(-\sqrt{3}\phi)} - 3B^2\exp{(-2\sqrt{3}\phi)}.
 \label{Vflatx2b}
\end{equation}
which is the first two terms in the superstring motivated perturbation
expansion, equation~(\ref{Vgenerala}), with $\gamma = \sqrt{3}$.

\subsection{Modified Powerlaw Inflation}

The next case we treat is superficially similar to the last, with
\begin{equation}
H(x) = Ax + Bx^2, \quad B>0 \label{Hxflat3}
\end{equation}
but the evolution we derive is markedly different. The corresponding
solution is
\begin{eqnarray}
V(x) &=& A^2(3-2\xi^2)x^2 + 2AB(3- 4\xi^2)x^3 + B^2(3-8\xi^2)x^4.
\label{Vxflat3a} \\
a(x) &=&
\ao\left[\frac{\xo^2(A+2Bx)}{x^2(A+2B\xo)}\right]^{1/4\xi^2}
\label{axflat3} \\
t(x) &=& \frac{1}{A\xi^2}\left[\frac{1}{2}
\left(\frac{1}{x}-\frac{1}{\xo}\right) -
\frac{B}{A}\log{\left(\frac{\xo(A+2Bx)}{x(A+2B\xo)}\right)}\right]
\label{txflat3}
\end{eqnarray}

For this choice of $H(x)$, $V(x) \rightarrow \infty$ as $x\rightarrow
\infty$ and $\xi \leq \sqrt{3/8}$. For other values of $\xi$, $V(x)$ is
negative for large $x$. When $\sqrt{3/8} < \xi < \sqrt{3/2}$ the
potential has a local maximum at
\begin{equation}
x_a = \frac{A}{B}\frac{2\xi^2 - 3}{3-8\xi^2}
\end{equation}
but for $\xi \geq \sqrt{3/2}$, $V(x)<0$ for all allowable values of $x$.
However, $a(t)$ is always increasing and there are no
solutions for which the universe reaches a maximum size and then
contracts back to a future singularity.  This solution starts from a
singularity though, since in the limit $x \rightarrow \infty$,
\begin{equation}
t = \frac{1}{A\xi^2}\left[
\frac{B}{A}\log{\left(\frac{A+2B\xo}{2B\xo}\right)} -
\frac{1}{2\xo}\right], \quad a = 0.
\end{equation}

In this instance, the logarithmic term in $a(x)$, never dominates and
the expansion is always approximately powerlaw. Because this exact
solution implicitly requires the value of $\dot{\phi}$ to be large when
it is in the region containing the local maximum, the conditions that
gave rise to a period of exponential expansion for the solution given by
equations~(\ref{axflat2}) and (\ref{txflat2}) are not be satisfied in
this case. Of course the possibility of a period of exponential
expansion in the general solution to the potential,
equation~(\ref{Vxflat3a}), is not ruled out.

For this solution
\begin{equation}
\epsilon(x) = 2\xi^2 \left( \frac{A + 2Bx}{A + Bx}\right)^2.
\end{equation}
If $\xi > \sqrt{1/2}$, $\epsilon > 1$ and there is no inflationary
phase. Alternatively, if $\xi < \sqrt{1/8}$ the solution is inflationary
at all times. For intermediate values of $\xi$, the inflationary era
will be preceeded by a period of powerlaw expansion, $a \propto t^p$,
but with $p< 1$.

Again, there are three special cases for which one of the terms in
$V(x)$ is zero, and a potential with two terms and independent
coefficients results.

\subsection{Exact solution for a superstring motivated potential}

Starting from $H = Ax - Bx^3$, we find the potential,
\begin{equation}
V(x) = A^2(3-2\xi^2)x^2 + 6AB(2\xi^2 -1)x^4 + 3B^2(1-6\xi^2)x^6,
\label{Vxflat4a}
\end{equation}
which has the form of the perturbation expansion suggested by
superstring theory, equation~(\ref{Vgenerala}), for all values of $\xi$.
\begin{eqnarray}
a(x) &=&
\ao\left[\frac{\xo^3(A-3Bx^2)}{x^3(A-3B\xo^2)}\right]^{1/6\xi^2}
\label{axflat4} \\
t(x) &=& \frac{1}{2A\xi^2}\left\{\frac{1}{x} - \frac{1}{\xo}  +
\frac{C}{2}\log{\left[\frac{(1-Cx)(1+C\xo)}{(1-C\xo)(1+Cx)}\right]}\right\}
\label{txflat4}
\end{eqnarray}
where $C = \sqrt{3B/A}$.

For all $\xi$ there is a local maximum at $x = 1/C$. For $\xo > 1/C$,
$x$ will increase as the universe evolves, and it will eventually reach
a maximum size and collapse towards a singularity.  For $\xo < 1/C$, the
expansion will continue indefinitely. If $\xo
\approx 1/C$ then $a(t)$ will initially be approximately exponential,
giving way to powerlaw behaviour at later times.

Again, there are three special values of $\xi$ for which one of the
terms drops out of the potential. In particular, when $\xi =
1/\sqrt{2}$, the potential is
\begin{equation}
V(\phi) = 2A^2\exp{(-\sqrt{2}\phi)} - 6B^2\exp{(-3\sqrt{2}\phi)}
\end{equation}
and $\gamma = \sqrt{2}$, the value derived from superstring theory,
although the term in $\exp{(-2\sqrt{2}\phi)}$ is missing.

Setting $H = Ax + Bx^3$ gives similar results to those found for
$H=A+Bx^2$. The potential and the scale factor are found simply by
changing the sign of $B$ in equations~(\ref{Vxflat4a}) and
(\ref{axflat4}). The time, $t(x)$ is
\begin{equation}
t(x) = \frac{1}{2A\xi^2}\left[\frac{1}{x} - \frac{1}{\xo}
+ C\left(\tan^{-1}{(x)} - \tan^{-1}{(\xo)}\right)\right]
\label{txflat5}
\end{equation}
This model with begins with a singularity, expands indefinitely and
can have a mixture of non-inflationary and powerlaw expansion, depending
on the value of $\xi$.

\section{Discussion}

We have found a number of new, exact scalar field cosmologies by
deriving the potential that produces a specified form of the density,
$\rho$, or the Hubble parameter, $H$. The technique used here typically
does not produce a complete solution for a given potential. In
particular, the solutions presented in this paper do not allow the
initial value of $\dot{\phi}$ to be chosen arbitrarily. However, it is
easy to apply and a large number of potentials can be examined and
interesting examples isolated, which can then be studied with more
complicated analytic or numerical methods.

In curved Robertson Walker spacetimes, we find a mixture of exact inflationary
and non-inflationary models. We generalise solutions previously
found by \OandT\ \cite{OzerET1992a} and develop analagous results for a
universe with negative spatial curvature.

In flat spacetime we have presented exact solutions that have several distinct
phases, distinguished by the properties of the scale factor, $a(t)$. As
well as models with a smooth transition between inflationary and
non-inflationary expansion, we have found instances where the
inflationary growth is a combination of exponential and powerlaw
expansion. We are not aware of any other exact scalar field cosmologies
that are of a comparable complexity and diversity. If a realistic model
possessed these properties,  the resulting spectrum of primordial
density perturbations would be comparatively complex, assuming that
there is not so much powerlaw growth that the fluctuations produced
during the era of exponential expansion are shifted beyond the present
horizon.

We have concentrated on solutions with relatively simple potentials
but the method used here can be used to generate an arbitrary number of
exact solutions where the potential consists of exponential terms. While
we can find only a handful of such solutions in curved spacetime, the
simple relationship between $H$ and $V(x)$ when $k=0$,
equation~(\ref{Vxflat}), means that large numbers of exact solutions can
be obtained in flat spacetime and we have not exhausted their
possibilities in this paper.

\begin{center}
\begin{figure}[]
\begin{center}
{\footnotesize
\begin{minipage}{120mm}
\epsfxsize=120mm
\leavevmode
\epsfbox{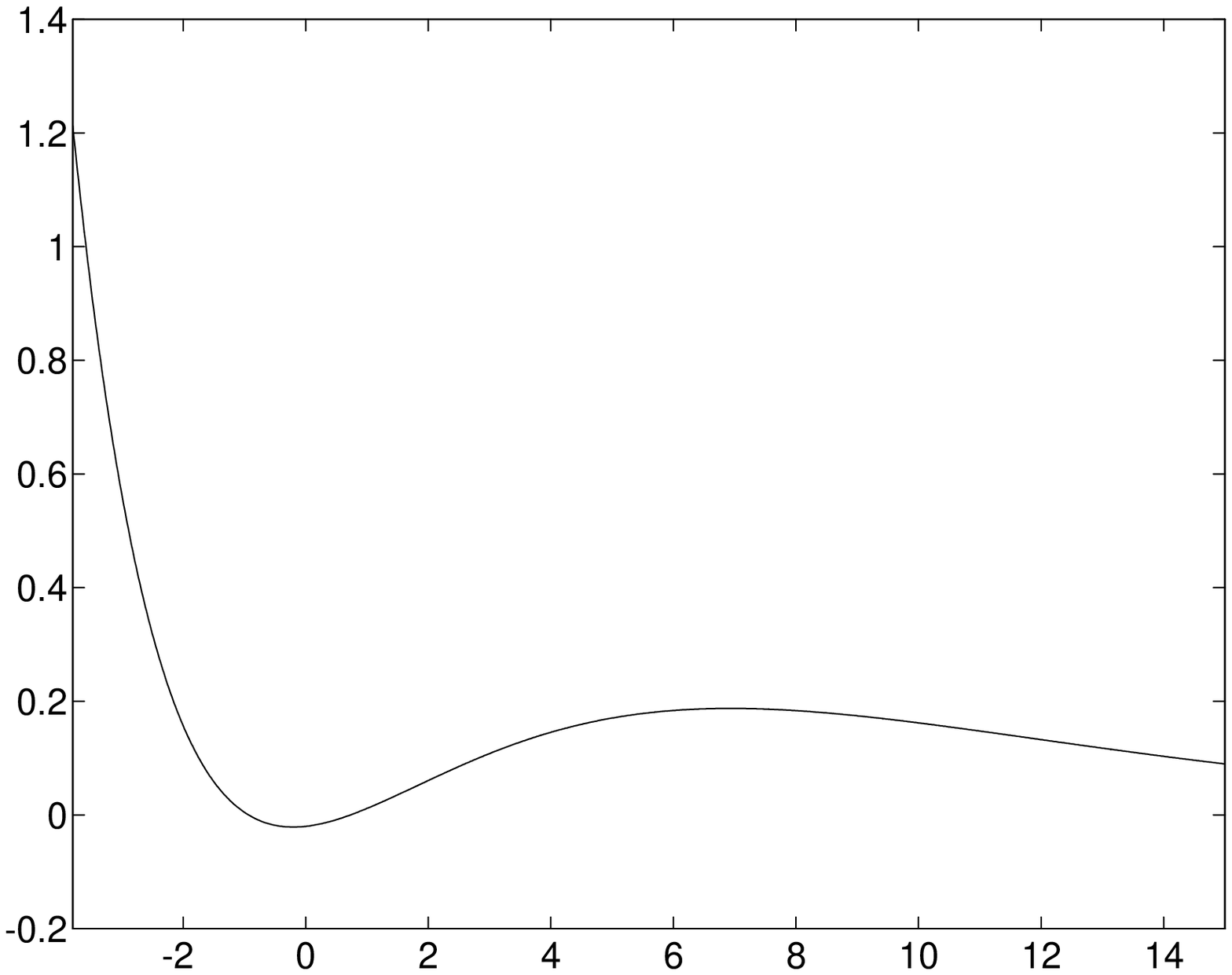}
\end{minipage}
} % end footnotesize
\caption[]{The potential, equation~(\ref{Vxflat2a}), is plotted against
$\phi$ for $\xi = 0.1$, $A=B=1$. When $x = e^{-\xi\phi} \approx x_a$
then the field is close to the local maximum of $V(\phi)$ and the scale
factor, $a(t)$ will be growing exponentially. As the universe evolves,
the field rolls away from the local maximum and the expansion makes a
smooth transition to powerlaw growth.}
\label{Vphi_fig}
\end{center}
\end{figure}
\begin{figure}[]
\begin{center}
{\footnotesize
\begin{minipage}{120mm}
\epsfxsize=120mm
\leavevmode
\epsfbox{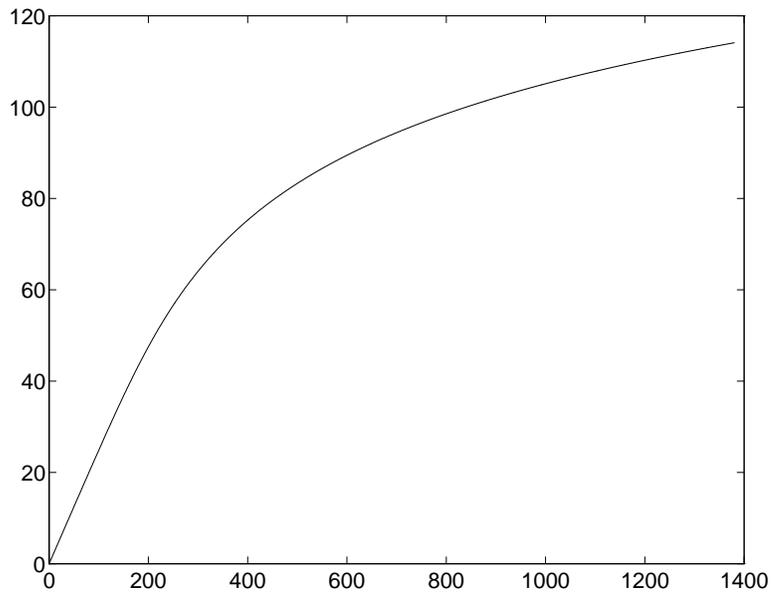}
\end{minipage}
} % end footnotesize
\caption[]{The scale factor, $\log{a(t)}$, is plotted against the time
$t$ for a model where the potential has the specific form illustrated in
Figure~(\ref{Vphi_fig}) and $\xo = 0.98x_a$. When $t \lesssim 200$,
$\log{a}$ is approximately proportional to the time, $t$ and the
universe is growing exponentially with powerlaw inflation taking place
at later times. At larger values of $\xi$ the amount of exponential
expansion is reduced, but the transition to powerlaw growth is more
abrupt. }
\label{logavt_fig}
\end{center}
\end{figure}
\end{center}

\end{document}